\documentclass[10pt,preprint]{aastex}
\usepackage{lscape}
\usepackage{natbib} 
\usepackage{amsmath}
\begin{document}

\newcommand{\teff}{\ifmmode T_{\rm eff} \else $T_{\mathrm{eff}}$\fi} 
\newcommand{\logg}{\ifmmode \log g \else $\log g$\fi} 
\newcommand{\mbol}{\ifmmode M_{\rm bol} \else $M_{\mathrm{bol}}$\fi} 
\newcommand{\lL}{\ifmmode \log \frac{L}{L_{\odot}} \else $\log \frac{L}{L_{\odot}}$\fi} 
\newcommand{\mdot}{$\dot{M}$} 
\newcommand{\myr}{M$_{\odot}$ yr$^{-1}$} 
\newcommand{\vsini}{$V$ sin$i$} 
\newcommand{\vinf}{$v_{\infty}$} 
\newcommand{\vturb}{v$_{turb}$} 
\newcommand{\vesc}{v$_{esc}$} 
\newcommand{\kms}{km s$^{-1}$} 
\newcommand{\msun}{\ifmmode M_{\odot} \else $M_{\odot}$\fi}

\title{Massive binaries in the vicinity of Sgr\,A*}
%\footnote{Based on observations collected at the ESO Very Large Telescope (programs 075.B-0547, 076.B-0259, 077.B-0503, 179.B-0261 and 183.B-0100)}} 

\author{
O.~Pfuhl\altaffilmark{a,}\altaffilmark{*},
T.~Alexander\altaffilmark{b},
S.~Gillessen\altaffilmark{a}, 
F.~Martins\altaffilmark{c},
R.~Genzel\altaffilmark{a,d},
F.~Eisenhauer\altaffilmark{a},
T.~K.~Fritz\altaffilmark{a} and
T.~Ott\altaffilmark{a}  
%N.~Blind\altaffilmark{a}
%A.~Sternberg\altaffilmark{b}
} 
 
%\email{martins@mpe.mpg.de} 
 
 
 \altaffiltext{a} {Max-Planck-Institut f\"ur Extraterrestrische Physik, 85748 Garching, Germany} 
% \altaffiltext{b} {Sackler School of Physics and Astronomy, Tel Aviv University, Israel} 
\altaffiltext{b}{Faculty of Physics, Weizmann Institute of Science, P.O. Box 26, Rehovot 76100, Israel}
 \altaffiltext{c} {LUPM, Universit\'e Montpelier 2, CNRS, Place Eug\`ene Bataillon, F-34095, Montpellier, France} 
 \altaffiltext{d} {Department of Physics, University of California, Berkeley, CA 94720, USA} 

% 
%  \altaffiltext{d} {Leiden University, Leiden Observatory and Lorentz Institute, , NL-2300 RA Leiden, the Netherlands} 
%  
%  \altaffiltext{e} {LESIA, Observatoire de Paris, CNRS, UPMC, Universit�� Paris Direrot, Meudon, France} 
%  \altaffiltext{f} {Faculty of Physics, Weizmann Institute of Science, Rehovot 76100, Israel} 
%  \altaffiltext{g} {Department of Physics \& Astronomy, University of Leicester, Leicester, UK} 
%  \altaffiltext{h} {Harvard-Smithsonian Center for Astrophysics, 60 Garden Street, Cambridge, USA} 

%  \altaffiltext{j} {IRAM Grenoble, 300 rue de la piscine, F-38406 Saint Martin d'Heres, France} 
 \altaffiltext{*} {correspondence: O.~Pfuhl, pfuhl@mpe.mpg.de}

\begin{abstract}
A long-term spectroscopic and photometric survey of the most luminous and massive stars in the vicinity of the super-massive black hole Sgr\,A* revealed two new binaries; a long-period Ofpe/WN9 binary, GCIRS\,16NE, with a modest eccentricity of 0.3 and a period of 224 days and an eclipsing Wolf-Rayet binary with a period of 2.3 days. Together with the already identified binary GCIRS\,16SW, there are now three confirmed OB/WR binaries in the inner 0.2\,pc of the Galactic Center. Using radial velocity change upper limits, we were able to constrain the spectroscopic binary fraction in the Galactic Center to $F_{\rm SB}=0.27^{+0.29}_{-0.19}$ at a confidence level of $95\%$, a massive binary fraction similar to that observed in dense clusters. The fraction of eclipsing binaries with photometric amplitudes $\Delta m>0.4$ is $F^{\rm GC}_{\rm EB}=3\pm2\%$, which is consistent with local OB star clusters ($F_{\rm EB}=1\%$). Overall the Galactic Center binary fraction seems to be close to the binary fraction in comparable young clusters.    
\end{abstract}


\maketitle

\section{Introduction}\label{sec:introduction}
The Milky Way nuclear star cluster (NSC) is the closest galactic nucleus and therefore target of detailed observations over the last few decades. It offers the unique possibility to resolve the stellar population and to study its composition and the dynamics close to a central black hole at an unrivaled level of detail. 
Precision astrometry of the innermost stars over almost two decades has proven the existence of a $4.3 \times 10^6 \msun$ supermassive black hole (SMBH) \citep{eisenhauer05,ghez08,gill09}.  \\
Past studies of the Milky Way's NSC found that the stellar population can be divided into two classes: the cool and evolved giant stars and the hot and young main-sequence$/$post-main sequence stars.  While the bulk of the stellar population is $\rm>5\,Gyrs$ old \citep{pfuhl11}, the existence of the massive young stars is evidence for very recent star formation \citep{forr87,allen90}. The most massive stars (WR$/$O stars) reside in a combination of a prominent warped disk, a second disk-like structure highly inclined relative to the main disk, and a more isotropic component \citep{paum06,lu09,bartko09} at a projected distance of 0.8\arcsec-12\arcsec~ from Sgr\,A* ($\rm 1\arcsec \equiv 0.04\,pc$, assuming a distance of $R_0=8.3$\,kpc). The GC disks must have formed in a rapid star burst $\sim 6\rm \,Myrs$ ago \citep{paum06,bartko10}, with a highly unusual initial mass function (IMF) that favored the formation of massive stars \citep{bartko10,2013ApJ...764..155L}. This extreme IMF deviates significantly from the standard Chabrier$/$Kroupa IMF with a powerlaw slope of $\alpha=-2.3$ \citep{kroupa01,chabrier03} and seems to exist only in the vicinity of the SMBH. 
The extreme IMF is currently pictured as the result of an infalling gas cloud that settled in a disk around the SMBH. Compressional heating of the fragmenting disk due to the tidal field of the SMBH raised the gas temperature, leading to the formation of massive stars \citep{2003ApJ...590L..33L,bonnell08}.
The fragmentation of an accretion disk however is not only expected to produce massive stars, but also to favor the formation of binary systems \citep{alexander08}. Fast cooling (shorter than the orbital timescale $\approx$ 1000 yrs) in fragmenting self-gravitational disks leads to a dramatic increase in the formation of binaries. The simulations predict a fraction of multiple systems close to unity in that case.

Apart from the massive O-star disks, a second population of ordinary B-stars can be found in the innermost 1\arcsec~ around the SMBH, the so called S-stars \citep{eisenhauer05,ghez08,gill09}. The origin of the S-stars is a mystery. In-situ formation seems impossible due to the strong tidal forces of the SMBH. On the other hand, inward migration from the greater GC area is limited by the short main-sequence lifetime of only a few ten Myrs. This requires the stars to have formed not far from todays location.
One of the currently favored mechanisms that explains the formation of the S-stars is a 3-body interaction of the SMBH and a binary system \citep{gould03,perets07}. The binary gets disrupted by the SMBH \citep[Hills' mechanism; ][]{1988Natur.331..687H}. The one companion is ejected and eventually ends as a hypervelocity star, while the other companion gets tightly bound to the SMBH.\\
Thus, the formation of both stellar populations is closely tied to the binarity of the massive O-stars in the Galactic Center. Although dynamical effects, like stellar capture or disruption due to stellar encounters, can change the initial binary fraction on timescales of only a few Myrs, the detection of binaries in the Galactic Center can constrain some of the formation models. 
\subsection{Observed binary fractions}\label{sec:observed_binary_fractions}
\textit{The binary fraction} of massive stars is subject of intense studies. The detection difficulties of long-period and extreme-mass ratio binaries makes it hard to estimate the true binary fraction and the underlying distribution of the binaries. However, it has been established that the binary fraction strongly varies with the environment. Dense stellar clusters seem to have lower binary fractions than less dense clusters. \cite{garcia01} tabulated binary fractions of massive O- and B-type stars for various clusters and concluded that the binary fraction decreases from 80\% to 14\% with increasing cluster density. The lowest fraction was found in Trumpler 14, one of the densest ($\rm \sim10^5\,\msun \,pc^{-3}$) young clusters in our galaxy.\\ \cite{sana12} found in six low-density open clusters a binary fraction of 70\%. \cite{mason09} found very similar values for O-stars in comparable clusters, yet lower fractions of 59\% and 43\% for field and runaway O-stars. They claim that field O-stars are ejected cluster stars and the lower binary fraction is related to the ejection mechanism (supernovae and close encounters).\\
Similar to the correlation for massive stars, \cite{milone08} found a strong anti-correlation between globular cluster masses and the corresponding binary fractions. 
\\
Two theoretical approaches try to explain the environment dependence of the binary fraction. The first focuses on the conditions in the star-forming cloud that regulate the formation of multiple systems \citep[e.g. ][]{sterzik03}.
The second approach argues for a universal initial binary population, that is altered by dynamical evolution, which depends on the environment. While in the first scenario, the multiplicity only depends on the density of the cluster, the second scenario predicts that the multiplicity also depends on age. For example, \cite{marks12} were able to reproduce numerically the observed densities and binary fractions of eight young clusters through dynamical evolution, starting from an universal initial binary population.   
\\
While the anti-correlation of density and binarity is well established, the age dependence is still an open debate. \cite{sollima07} found a slight binary-age correlation in globular clusters. However a subsequent study of open clusters \citep{sollima10} remained inconclusive in this respect.\\
 
Binaries among low-mass stars ($\approx 1\,\msun$) seem to be less frequent than among massive stars. Recent studies suggest binary fractions between 4\% and 15\% \citep{sollima07,sommariva09} in old globular clusters. In less dense open clusters, \cite{geller12} found 29\% low-mass binaries. For comparison, the field binary fraction is 57\% \citep{duq91}, i.e. significantly enhanced compared to dense clusters.
\subsection{Binary statistics} \label{sec:binary_statistics}
Studies of large OB associations, such as Cygnus OB2, with hundreds of stars allowed to derive binary distribution statistics, namely binary mass ratio-, orbital separation- and eccentricity distributions. The observed distribution functions are found to be well described by power laws. \cite{kobul07} and \cite{kiminki12}, for example, found a mass ratio $q=M_{\rm sec}/M_{\rm prim}$ distribution $f(q)\propto q^\alpha$, with $\alpha=0.1 \pm 0.5$, a log-period distribution of $f({\rm log} P)\propto {\rm log} P^\beta$, with $\beta=0.2\pm0.4$ and an eccentricity distribution of $f(e)\propto e^\gamma$, with $\gamma=-0.6\pm0.3$.
A similar study of the Tarantula Nebula by \cite{sana12b} found a somewhat steeper mass ratio $\alpha=-1 \pm 0.4$ but also shorter periods $\beta=-0.45\pm0.3$.

\subsection{What is known about binaries in the Galactic Center}\label{sec:GCbinaries}
Due to the large distance of the GC and the extreme extinction in the optical \citep[$A_V > 30$; e.g.][]{fritz11}, the study of GC binaries is limited to the most massive early-type stars. \\
There is only one confirmed binary so far:

\begin{itemize}

\item IRS\,16SW consists of two equal 50\,\msun ~constituents, with a period of 19.5 days \citep{ott99,martins07}. The star is an eclipsing contact binary, which shows a large photometric and spectroscopic variability during its revolution.
\end{itemize}
However a few more stars were speculated to be binaries:
\begin{itemize}
\item  The bow-shock star IRS\,8, about 30\arcsec~ north of SgrA* was speculated to be a binary due to its apparently young age of only 3\,Myrs \citep{geballe06}. No binary signature has been detected so far, however the seemingly young age might be explained by the influence of a close companion on the primary evolution. We did not consider this star in our study due to its relatively large distance from the cluster center. Considering the steep radial profile of the early-type stars, it is not clear if IRS\,8 can be associated with the WR disk formation.

\item For IRS\,7E2, a massive Ofpe/WN9, \cite{paumard01} reported a significant radial velocity change with respect to a previous measurement by \cite{genzel00}. Unfortunately, we obtained only four additional epochs for that star. Among the few observations, we did not detected a significant radial velocity change. Yet, the star features very broad emission lines (FWHM=1140\,km/s), which show some intrinsic variability. Thus to conclude on a binarity of the star, more observations are required. So far it can only be considered a binary candidate.

\item Photometric variability of IRS\,29N was interpreted by \cite{rafelski07} as the potential signature of a wind colliding binary. However, older data from \cite{ott99} showed less variability and no periodicity. The star is classified as a dust producing WC star \citep{paum06}. Some irregular variability could thus be attributed to circumstellar dust. The stellar spectrum is very red and shows some extremely broad emission features. The width and the intrinsic variability of the features prevents a precise radial velocity measurement. Therefore we were not able to confirm or rule out a binarity of IRS\,29N.   

\item \cite{peebles07} classified a photometrically variable star with a period of $\sim42$ days as an eclipsing binary candidate. As for IRS\,8 the star has relatively large distance ($\approx27\arcsec$) from the cluster center and is therefore likely not related to the WR disk. Due to the large distance it was not included in our photometric and spectroscopic survey. However its spectroscopic confirmation is a viable target for future observations.  

\end{itemize}

%
%The paper is organized as follows; In section \ref{sec:obs_and_proc} we present the data and in section \ref{sec:spec_class} the spectral classification and calibration of spectral indices. In section \ref{sec:const_HRdiag} we construct the H-R diagram that is then fit with model populations in section \ref{sec:SFR_fit}. The star formation history, IMF and mass composition of the nuclear cluster is presented in section \ref{sec:results}. The results are discussed and compared with other works in section \ref{sec:discussion}. We conclude in section \ref{sec:conclusion}.

\section{Observations and data processing}\label{sec:obs_and_proc}
This work relies on spectroscopic and imaging data obtained at the VLT in Cerro Paranal Chile between 2003 and 2013. The observations were carried out under the program-ids 075.B-0547, 076.B-0259, 077.B-0503, 087.B-0117, 087.B-0280, 088.B-0308, 288.B-5040, 179.B-0261 and 183.B-0100.
 \subsection{Imaging and photometry}
 The photometric data were obtained with the adaptive optics camera NACO \citep{rousset03,2003SPIE.4841..944L}. The photometric reference images were taken on the 29th of April 2006 and on the 31st of March 2010. We used the $H$- and $Ks$-band filter together with a pixel scale of 13\,$\rm{mas/pixel}$. To each image we applied sky-subtraction, bad-pixel and flat-field correction \citep{trippe08}. All images of good quality obtained during the same night were then combined to a mosaic with a field of view of $\approx20\arcsec \times20\arcsec$. In total we used 102 $Ks$-band images and 34 $H$-band images with temporal spacings between a few hours and up to 9 years to construct lightcurves for a few thousand stars within the FoV. 
\subsection{Spectroscopy}
Our spectroscopic data were obtained with the adaptive optics assisted integral field spectrograph SINFONI \citep{eis03, bon04}. In total we used 45 observations obtained between spring 2003 and summer 2013 with pixel scales between $50 \times 100$ and $125 \times 250$\,mas. The data output of SINFONI consists of cubes with two spatial axes and one spectral axis. Depending on the plate scale, an individual cube covers $3.2\arcsec \times3.2\arcsec$ or $8\arcsec \times8\arcsec$; the spectral resolution varies between 2000 and 4000 depending on the chosen bandpass and the field-of-view. We used the data reduction SPRED \citep{schreiber04,abu06}, including bad-pixel correction, flat-fielding and sky subtraction. The wavelength scale was calibrated with emission line gas lamps and fine-tuned on the atmospheric OH lines. Finally we removed the atmospheric absorption features by dividing the spectra through a telluric spectrum obtained in each respective night.

\subsection{Spectroscopic sample selection}
Of order 200 early-type stars are known within 1\,pc from Sgr A*. Their spectral types range from the most luminous O/WR stars with main-sequence lifetimes of a few Myrs to early B-stars with main-sequence lifetimes of several 10\,Myrs. Stars fainter than B-dwarfs ($m_K>16$) are too faint to be identified spectroscopically. Among the known early-type stars we chose the brightest ones ($m_K <12$) with prominent emission or absorption lines, that allowed a precise radial velocity measurement. We excluded stars with fit errors larger than 20\,km/s. For instance, we excluded several bright WR stars with very broad wind emission lines. Our final sample consisted of 13 stars in close proximity to Sgr A*, which we repeatedly observed with SINFONI. Additionally we used archival data from the same region, that gave us up to 45 independent observations per star (see table\,\ref{tab:data}). The observations cover period spacings between one day up to 9 years.
\subsection{Velocity measurement and uncertainty} \label{sec:vel_uncertainty}
For the velocity measurement, we chose the most prominent spectral feature of each star. Depending on the spectral type, this was either the He\,I line at 2.059\,$\rm \mu m$, the He\,I line at 2.113\,$\rm \mu m$ or the Br$\gamma$ line at 2.166\,$\rm \mu m$. The absolute velocity of the individual stars was measured by fitting a Gaussian. In order to detect relative velocity changes, we cross-correlated each spectrum with a reference spectrum of the same star. This provided significantly better velocity residuals and scatter than individual Gaussian fits. All velocities were corrected for the earth bary- and heliocentric motion. The velocity uncertainties of the individual measurements are a combination of the fit errors, systematic errors (e.g. wavelength calibration) and intrinsic line variations of the stars. The formal fit errors in the best cases were as small as $\approx2$\,km/s. However, this does not include systemic errors like wavelength calibration and drifts of the spectrograph. To determine the overall velocity uncertainty (including systematics), we used three late-type stars contained in the integral field unit (IFU) fields and measured their velocities. Late-type giants are good spectroscopic references because they can't have close companions (due to their physical size of few AU) and they show only slow pulsations. Thus intrinsically their radial velocities are thought to be very stable. Observationally, they are well suited due to their prominent absorption features in the K-band, the CO bandheads. The absorption shape allows a very precise radial velocity measurement (fit errors $<2$\,km/s). It turns out that the late-type giants show a radial velocity scatter in our measurements of $\sim6\,$km/s RMS. We therefore estimate, that our systematic errors are of that order.   

\section{The long-period binary GCIRS\,16NE}\label{sec:16NE}
The star IRS\,16NE is the most luminous early-type star in the Galactic Center ($m_H\approx11.0$, $m_K\approx9.2$). With a bolometric luminosity of $L\approx 2.2\times 10^6\,L_{\odot}$ \citep{najarro97,martins07} it is even one of the most luminous stars in the Milky Way.
It is part of a young and massive population in the GC, thought
to be luminous blue variables \citep[LBVs, e.g. ][]{paumard01}. Of the same type are the stars IRS\,16C,
IRS\,16NW, IRS\,16SW, IRS\,33E, and IRS\,34W. IRS\,16NE is at least 0.5 magnitudes brighter than the other LBV stars \citep{paum06}. LBVs are evolved massive stars that populate a region in the H-R diagram, where the luminosity approaches the Eddington luminosity, which leads to instabilities in their atmospheres. Therefore those stars show strong variability in photometry and spectroscopy (Humphreys \& Davidson 1994). Characteristic for this stellar phase is strong wind-driven mass loss and drastic changes in the stellar temperature and radius. Given that strong outbursts have not been observed yet for the six stars, they are thought to be LBVs in a stable phase \citep{martins07}. \\
\cite{martins06,tanner06,zhu08} recognized a significant radial velocity change of IRS\,16NE and speculated about a binary origin. However they were not able to deduce an orbital solution and deemed the star only a candidate.\\
After collecting another 6 years worth of data and effectively doubling the number of observations, we can finally confirm the binarity of IRS\,16NE. 

\subsection{Orbital solution and physical parameters}
\begin{figure}[!ht]
\begin{center}
\includegraphics[width=0.9\columnwidth]{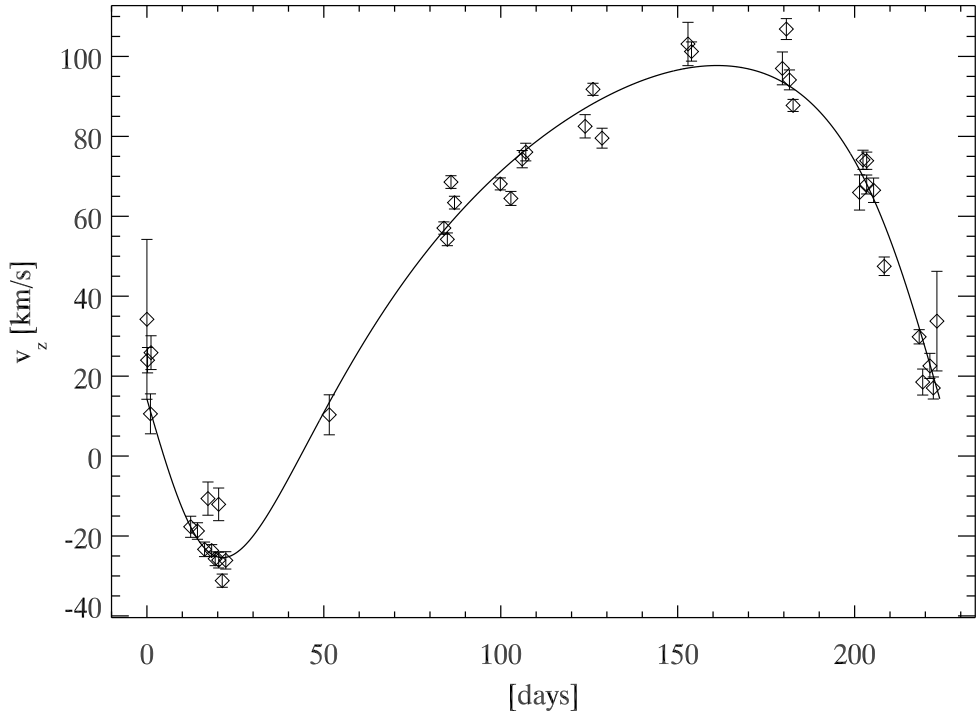}

\caption{Radial velocity curve of IRS\,16NE together with the best orbital
solution (for $P=224.09$ days and $e =0.32$, solid line). The typical uncertainty on the radial
velocity is $\pm 6.6\,{\rm km\,s^{-1}}$. Parameters for the best-fit solution are given in Table\,\ref{tab:parameter}.}
\label{fig:fold_period}
\end{center}
\end{figure}

\begin{figure}[!h]
\begin{center}
\includegraphics[width=0.5\columnwidth]{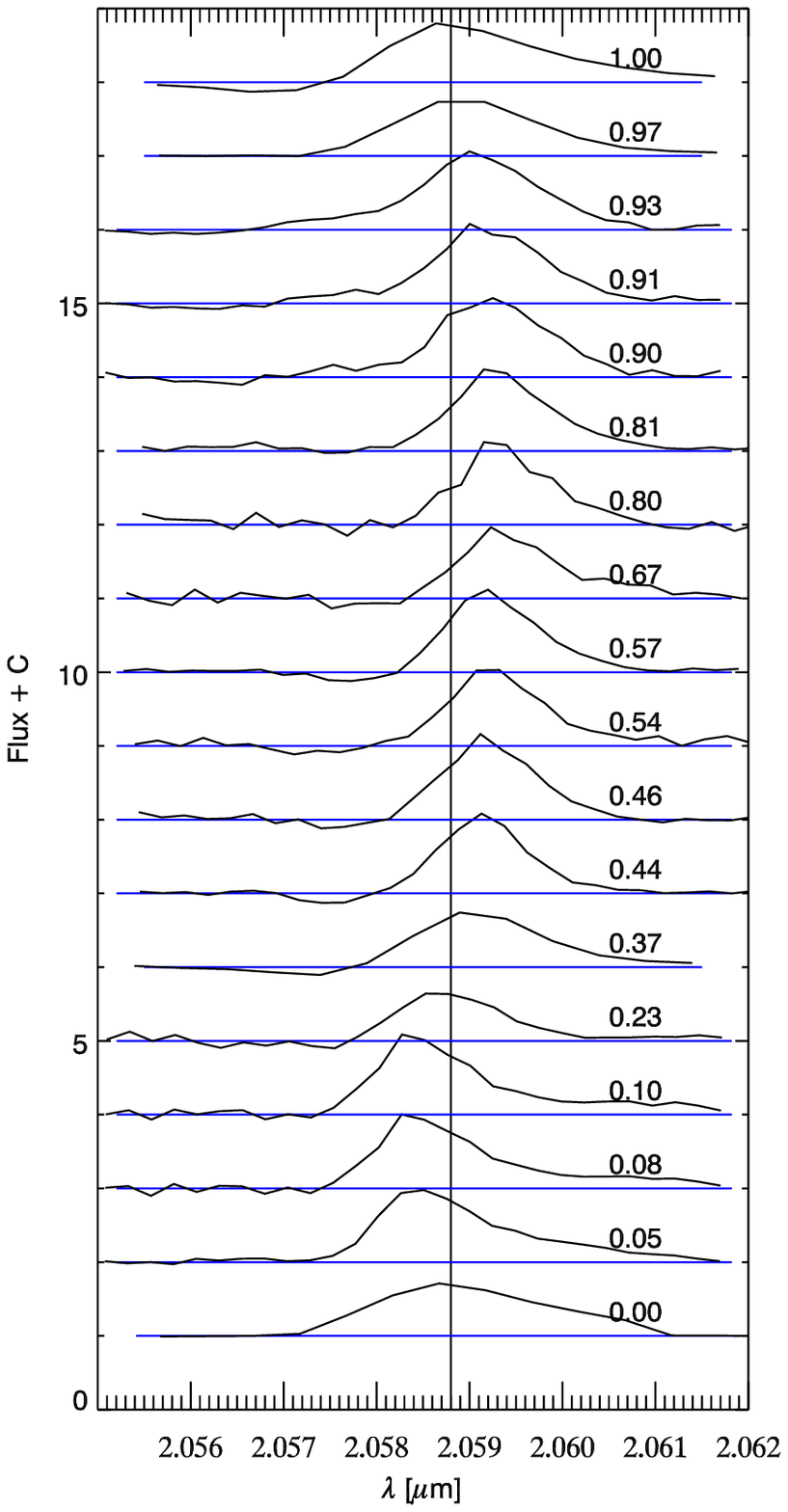}
\caption{Sequence of spectra following one orbital period of IRS\,16NE. The number indicates the period. The spectra have been corrected for the earth bary-centric motion. The solid line indicates the rest-wavelength of He\,I (2.059\,$\rm \mu m$)}
\label{fig:sequence}
\end{center}
\end{figure}

We obtained 43 spectra of IRS\,16NE, spread over roughly 10 years with spacings of a few days up to years.  
The orbit of a single line spectroscopic binary is defined by the period $P$, the eccentricity $e$, the systemic velocity $\gamma$, the longitude of periastron $\omega$, the time of periastron $T$ and the mass function $f(m)$. However, fitting periodic functions such as a binary signature can be problematic, especially if one tries to fit the period. Often the algorithm fails to converge or gets trapped in a local minimum. Therefore, we set the period as a prior and tried to fit the velocity curve for the given prior. We repeated the fitting for periods between 0.5 days up to 1000\,days with a spacing of 1\,day. The solution with the lowest $\chi^2$ of all individual fittings was taken as the true orbital solution. We used the IDL fitting tool \textit{MPFIT} \citep{2009ASPC..411..251M}, which provided a fast fitting algorithm, and which allowed defining parameter constrains. For stability reasons, we constrained the eccentricities $e<0.98$. Although the period was set as a prior, we allowed the fitting routine to adjust the period by $\pm0.5$\, days with respect to the prior period. Thus we could refine the crude 1\,day period sampling.\\
It turned out that the fitting clearly favored a solution with a period of 224.09\,days. The second best period had a factor four higher $\chi^2$ than the best solution. A folded sequence of spectra covering one orbital period is shown in Figure\,\ref{fig:sequence}. The orbit is clearly eccentric with $e=0.32$. Figure\,\ref{fig:fold_period} shows the folded radial velocity data together with the best-fit orbital solution. Since we were only able to measure the velocity of one companion, only the mass function;
\begin{equation}
f(m)=\frac{(M_2~{\rm sin}\,i)^3}{(M_1 +M_2)^2}=4.58\pm0.17 \,\msun \nonumber
\end{equation}
  of the system could be determined ($M_1$ is the mass of the observed star and $M_2$ is the mass of the unobserved star). However the spectroscopic similarity of the primary star with the known eclipsing binary IRS\,16SW argues for a primary mass close to 50\,\msun. This means that the companion mass is $\ge30\,\msun$. In case of an edge-on orbit ($i=90^\circ$) the secondary mass is 30\,\msun. A more massive companion requires a lower inclination. The lowest possible inclination is $\approx46^\circ$, for an equal mass companion. In fact, a roughly equal mass companion would explain the $\approx0.5$\,mag excess of IRS\,16NE compared to the other IRS\,16 stars. %Given that the secondary companion is not visible in the spectrum, it is most likely less massive than the primary, i.e we see a close to edge-on orbit. Yet, this does not mean that the companion's initial mass was lower than the primary mass. More evolved (i.e. initially more massive) WR stars have significantly higher mass-loss rates and expel a large fraction of their mass during the WR stage.\\
The binary IRS\,16NE, with a semi-major axis of $a\, {\rm sin}\, i =144\,{\rm \mu as}$ will be a valuable test-case for the upcoming 2nd generation VLTI instrument GRAVITY. With its unprecedented $\approx10\,{\rm \mu as}$ astrometric accuracy, it will be possible to determine the full orbital parameters of the system in less than one year of observations.
    
\begin{deluxetable}{lc}
\tablecaption{Orbital Parameters as Derived from the Analysis
of the Radial Velocity Curve\label{tab:parameter}}
\tablewidth{0pt}
\tablehead{
\colhead{Parameter} &
\colhead{Value}
}
\startdata
Semi-major axis, $a\,{\rm sin}\,i~ ({\rm 10^6\,km})$ & $179.1\pm7.3$ \\ 
Eccentricity, $e$ & $0.32\pm0.01$ \\
Systemic velocity, $v_0$ ($\rm km/s$)& $52.45\pm0.46$ \\
Semi-amplitude $K1$ ($\rm km/s$) & $61.57\pm1.7$ \\
Longitude of periastron, $\omega$ (deg) & $144.54\pm1.65$ \\
Orbital period, $P$ (days) & $224.09\pm0.09$\\
Time of periastron, $T$ (mjd)& $52523.63\pm1.47$\\
$f(m)$ ($M_{\odot}$) & $4.58\pm0.17$\\
\enddata
\end{deluxetable}
\section{Eclipsing binaries}\label{sec:eclipsing_binaries}
In order to find eclipsing binaries, we focused on the spectroscopically confirmed sample of early-type stars in the inner 10\arcsec~ around SgrA*. Among those stars, we selected those with average photometric errors $<0.1$ magnitudes. This left us with 113 early-type stars within the FoV of the NACO 13\,mas/pix camera. We checked each of the lightcurves for variability. About half of the stars showed non-periodic variability on a few 0.1 mag level as expected for OB stars \citep{lefevre09}. In order to detect periodic variability, we used the phase dispersion minimization \citep{stellingwerf78}, a widely used technique in Cepheid and eclipsing binary searches. The inspection of the individual periodograms allowed us to identify two periodic variables with short periods ($<100$\,days). The previously reported eclipsing binary IRS\,16SW \citep{ott99,martins07} with a period of 19.447 days and a new periodic variable with a period of 2.276 days. 
\subsection{The eclipsing binary E60}
\begin{figure}[!h]
\begin{center}
\includegraphics[width=0.9\columnwidth]{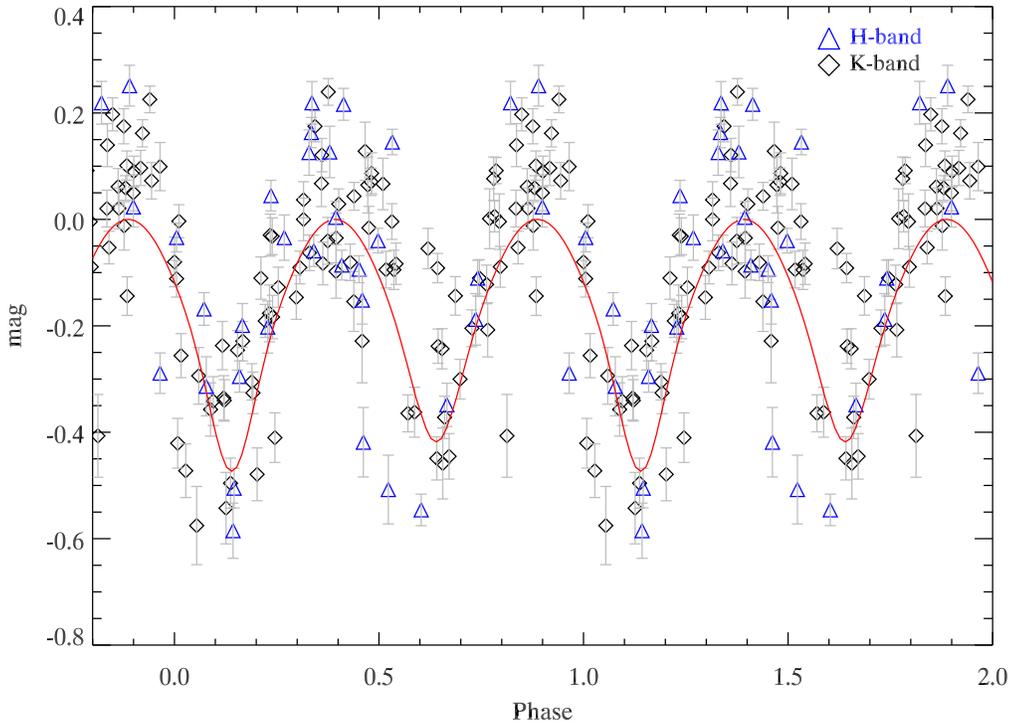}
\caption{Back-folded $H$- and $K$-band lightcurve of the eclipsing binary E60. The orbital period is 2.276 days. Overplotted is a model lightcurve calculated with the free program NIGHTFALL (for the parameters, see Table\,\ref{tab:NIGHTFALL_para}).}
\label{fig:lightcurve}
\end{center}
\end{figure}

\begin{figure}[!h]
\begin{center}
\includegraphics[width=0.9\columnwidth]{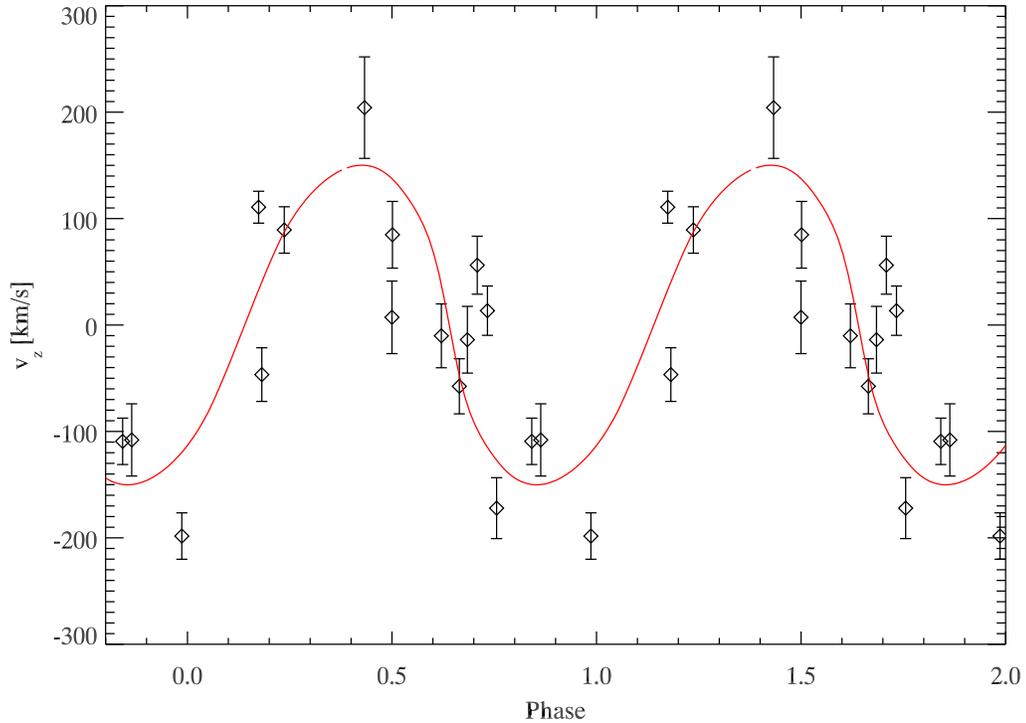}
\caption{Measured radial velocities (the systemic velocity is subtracted) of the eclipsing binary E60. The broad wind lines of the star leads to relatively large velocity errors. Overplotted is the model radial velocity, which was calculated using the same model as in Fig.\,\ref{fig:lightcurve}}
\label{fig:radialvel_E60}
\end{center}
\end{figure}

The new periodic star is the second reported eclipsing binary in the GC.  \cite{paum06} identified the star as a WN7 Wolf-Rayet type with $m_K=12.4$, located at $\Delta \alpha=-4.36\arcsec$ and $\Delta \delta =-1.65\arcsec$ from Sgr A*. Following the nomenclature of \cite{paum06}, the star is referred to as E60. The back-folded $H$- and $K$-band lightcurve can be seen in Fig.\,\ref{fig:lightcurve}. The color independent variability argues for an occultation event such as an eclipse of a companion. Variability due to pulsation or extinction from circumstellar dust typically leads to strong color changes. The WN7 star features broad emission lines, which results in relatively large radial velocity errors. Nonetheless, E60 shows a significant radial velocity change within days (only one companion is detectable). 
The radial velocity change is co-phased with the photometric periodicity, as the back-folded radial velocity curve indicates (Fig.\,\ref{fig:radialvel_E60}). Using the available photometric and spectroscopic data, we tried to model the binary with the program NIGHTFALL\footnote{The software was developed by R. Wichmann and is freely available at http://www.hs.uni-hamburg.de/DE/Ins/Per/Wichmann/Nightfall.html}.
The near sinusoidal lightcurve argues for a very close binary. In fact, to model the light curve, we had to use a Roche lobe filling factor of 1.1. This means that the companions are in contact. This is not surprising, given the short orbital period of only 2.276 days. Furthermore, the large photometric amplitude requires the inclination to be $>60^\circ$. The mass and the mass ratio of the system are essentially determined by the radial velocity amplitude. Unfortunately the few velocity measurements and the relatively large errors limit the ability to constrain those parameters. For a well determined fit, we would require more spectroscopic epochs, especially with short time spacings of only a few hours. However, we found a reasonable solution (see Table\,\ref{tab:NIGHTFALL_para}), that can reproduce the observations. 
\begin{center} 
\begin{table}[!h] 
\begin{center} 
\caption{Orbital parameters of the eclipsing binary E60 \label{tab:NIGHTFALL_para}} 
\begin{tabular}{lc} 
\hline\hline 
Parameter & Value\\ 
\hline    
Separation, $a\,~ (R_{\odot})$ & $22.6\pm3$ \\ 
Eccentricity, $e$ & $\approx 0$ \\
Systemic velocity, $v_0$ ($\rm km/s$)& $467\pm10$ \\
Semi-amplitude $K1$ ($\rm km/s$) & $150\pm7$ \\
Inclination, $i$ (deg) & $70\pm10$ \\
Orbital period, $P$ (days) & $2.276$\\
Mass ratio, $m$ & $2\pm0.5$\\
$M_{\rm system}$ ($M_{\odot}$) & $30\pm10$\\
\hline 
\end{tabular} 
\end{center} 
\end{table} 
\end{center}
We modeled the system with a total system mass of 30\,\msun~ and a mass ratio of two. An uneven mass ratio is required due to the relatively low velocity change for the given orbital period and inclination. Thus the primary mass is 20\,\msun~ and the secondary mass is 10\,\msun. In fact, those masses are typical for evolved WR stars of similar brightness and spectral type WN7 \citep[compare Table\,2 in][]{martins07}. The stellar radii of WN7 stars inferred by \cite{martins07} are between 10\,$R_{\odot}$ and 18\,$R_{\odot}$, which matches the inferred binary contact separation of $22.6\,R_{\odot}$.  \\
The binary E60 has a remarkably high systemic radial velocity of $422\pm10$\,km/s. The proper motion of the system is $4.73\pm0.14$\,mas/yr (Fritz, priv. comm.), which corresponds to $184\pm6$\,km/s at the distance of the GC. Thus the total systemic velocity is $v_{\rm 3D}=460\pm12\, \rm km/s$. The velocity exceeds (2\,$\sigma$) the escape velocity ($v_{\rm esc}\approx440\,\rm km/s$) at 4.7\arcsec~ projected distance from Sgr A*. The actual 3D distance of E60 could be larger, i.e. the escape velocity could be even lower. On the other hand, the absolute velocity of the star is probably less certain than the formal fit error might indicate. In particular the strong wind lines of E60 could be biased by the actual wind morphology. In any case, the star seems to be at best marginally bound to Sgr A*.
\section{Determining the spectroscopic binary fraction}\label{sec:detection_probability}
The observed spectroscopic binary fraction (two out of 13 stars; IRS\,16SW and IRS\,16NE), represents only a lower limit to the true binary fraction. 
Note, the eclipsing binary E60 was not included in the initial spectroscopic sample because its radial velocity uncertainty did not match the criterion. It was detected later due to its photometric variability. Only after the photometric detection, the radial velocity change was found.
In order to keep the spectroscopic sample unbiased, E60 therefore is not considered in the spectroscopic binary fraction.\\
Naturally, the probability to detect a stellar companion depends on the primary mass, the secondary mass (i.e. the mass ratio $q$), the eccentricity $e$ and the orbital period $P$. It also depends on the number of observations, the radial velocity uncertainty and how well the orbital period is sampled. To derive the true binary fraction, it is therefore necessary to take the detection incompleteness into account.   
\begin{landscape}

\begin{center} 
\begin{table*} 
\begin{center} 
\caption{Bright early-type stars targeted in the spectroscopic survey \label{tab:data}} 
\begin{tabular}{lcccccccc} 
\hline\hline 
id & sp type\tablenotemark{a} & N data\tablenotemark{b}& $v_{\rm z}$ [km/s]  & RMS($v_{\rm z}$)\tablenotemark{c} [km/s]& Fit error [km/s] & $M [\msun]$\tablenotemark{d} & $p_{\rm det} (\rm Kiminki)$\tablenotemark{e} & $p_{\rm det}(\rm Sana)$\tablenotemark{f}\\ 

\hline    
IRS\,16SW\tablenotemark{g} &Ofpe/WN9 & 25 & 459.5  & -  & 20 & 50 & Binary & Binary  \\
\hline 
IRS\,16NE & Ofpe/WN9 & 43 & 52.5  & 46.4  & 2.2 & $>40$ & Binary & Binary  \\
IRS\,16C & Ofpe/WN9& 43 & 186  & 10.3  & 2.7 & 40 & 0.70 & 0.71 \\
IRS\,16NW &Ofpe/WN9 & 37 & 17  & 11.4  & 2.7 & 40 & 0.70 & 0.71 \\
IRS\,33E &Ofpe/WN9 & 42 & 214  & 10.1  & 3.6 & 40 & 0.73 & 0.74 \\
IRS\,34W & Ofpe/WN9& 19 & -184  & 6.5  & 2.3 & 25 & 0.76 & 0.78 \\
IRS\,13E2 & WN8& 23 & -2  & 20.5  & 5.5 & 82.5 & 0.59 & 0.61 \\
IRS\,16SE2 & WN5/6& 38 & 191  & 15.4  & 6.5 & 17.2 & 0.52 & 0.55 \\
IRS\,29NE1 & WC8/9& 27 & -99  & 27.5  & 19.8 & 25 & 0.36 & 0.41 \\
%IRS\,16SE1 & WC8/9& 25 &  - & 35.9  & 26.3 & 20 & - & - \\
IRS\,33N (64) &B0.5–1 I & 31 &  93 & 22.5  & 7.3 & 25 & 0.43 & 0.47 \\
IRS\,16CC & O9.5–B0.5 I & 35 & 145  & 27.1  & 8.5 & 40 & 0.42 & 0.47 \\
IRS\,16S (30)& B0.5–1 I & 30 & 123  & 15.0  & 7.3 & 40& 0.63 & 0.65 \\
IRS\,1E & B1-3 I& 14 & 18  & 19.6  & 8.7 & 25 & 0.46 & 0.51 \\
\hline 
\end{tabular} 
\footnotetext{Spectral type taken from \cite{paum06}.}
\footnotetext{Number of individual spectroscopic observations.}
\footnotetext{Standard deviation of the N individual velocity measurements.}
\footnotetext{Main-sequence masses are derived from the H-R diagrams of \cite{paum06} and \cite{martins07}.}
\footnotetext{Companion detection probability determined with the distribution functions from \cite{kiminki12} (see Sec.\,\ref{sec:binary_statistics}).}
\footnotetext{Companion detection probability determined with the distribution functions from \cite{sana12b} (see Sec.\,\ref{sec:binary_statistics}).}
\footnotetext{Binary published in \cite{martins06}. }

\end{center} 
\end{table*} 
\end{center}
\end{landscape}
\subsection{Companion detection probability}\label{sec:detection_probability}
Thanks to the long term monitoring of the stars in our sample, we were able to set tight constrains on radial velocity changes of the respective stars. The masses of the stars can be quite well constrained from their luminosities and temperatures, i.e. the positions in the H-R diagram \citep[e.g.][]{paum06,martins07}.
In order to determine the binary detection completeness, we used the assumption that the binaries in the Galactic Center follow similar distribution functions as galactic and extra-galactic OB clusters (described in Section\,\ref{sec:binary_statistics}). This assumption might seem somewhat arbitrary, since it is not obvious that the distribution functions in disk star forming regions are applicable to the special environment around a massive black hole. 
For lack of better alternatives and keeping this limitation in mind, we used the observed distribution functions for a Monte-Carlo analysis. \\
For each star in the observed sample we created $10^6$ artificial companions, where the mass ratio $0.1<q<1$, the eccentricity $0<e<0.9$ and the period $1<P<1000$ were drawn with the observed distribution functions (Sec.\,\ref{sec:binary_statistics}). The longitude $\omega$, the time of periastron $T$ and the system inclination (${\rm cos}\,i$) were drawn with uniform probability. Each companion realization resulted in an artificial radial velocity curve, which was sampled with the same time spacing as the actual observations. From the artificial discrete radial velocity points, a velocity RMS was calculated. To account for the measurement uncertainties, we added in squares a systematic velocity uncertainty of 6\,km/s (see Sec.\,\ref{sec:vel_uncertainty}).\\
 The ratio of companion realizations with a velocity RMS equal or greater than the observed RMS, to the total number of realizations was then taken as the probability of detecting companions. The Monte-Carlo simulation did not produce false positive detections, i.e. we assumed the false positive detection probability to be zero. This approach seems to be justified because we only deemed stars as binaries were a unique orbital solution could be found. 
Table\,\ref{tab:data} states the detection probabilities for the stars in our sample. To check the robustness of the results, we ran the simulations with the two observed binary distribution functions of \cite{sana12b} and \cite{kiminki12}. Although the distribution functions seem quite different, the detection probabilities for both cases turned out to be very similar. The Sana distribution features on-average lower companion masses compared to the Kiminki study but also on-average shorter periods. This causes the results to be almost identical. The stars with the lowest radial velocity RMS have detection probabilities $>$ 0.7. In other words, the chance to have missed a companion is lower than 0.3. Naturally the detection probability can never reach unity simply due to the random inclination. Systems observed close to face-on are not detectable. The sample stars with the largest radial velocity errors, low primary masses or only few observations have detection probabilities as low as 0.36. 
\subsection{Spectroscopic binary fraction}
Assuming that all our sample stars intrinsically have the same probability to have a companion (the stars are all massive OB/WR stars), it is reasonable to use the average detection probability $\left\langle p_{\mathrm{det}}\right\rangle =0.57$ (Table 3). The observed binary fraction (2 binaries out of 13 sources) is $F_{\mathrm{obs}}=2/13\approx0.15$, which after the correction for $p_{\mathrm{det}}<1$ rises to $F_{SB}=F_{\mathrm{obs}}/p_{\mathrm{det}}\approx0.27$. The distribution of binaries among the observed sources can be viewed as the outcome of a binomial process with probability $P_{SB}$, which is determined by the physics of binary formation. The detection-corrected binary fraction $F_{SB}$ is then the sample estimator of $P_{SB}$.
The confidence interval around it can be obtained in the small sample size limit by the Wilson score interval with continuity correction \citep{newcombe98,brown01}. We thus find the 95\% confidence
interval to be $P_{SB}\in[0.08,0.56]$, or $F_{SB}=0.27_{-0.19}^{+0.29}$.
The lower bound of the binary fraction is lower than the observed fraction. This takes into account that we could have been 'lucky' in our choice of targets.
While the uncertainties are quite large, we can exclude a binary fraction
close to unity at high confidence. For example, $P_{SB}>0.85$ is
ruled out at the $99.999999\%$ level.
\section{Eclipsing binary fraction}
Estimating the true eclipsing binary fraction in the Galactic Center is non-trivial because the detection probability depends strongly on the data sampling, the duration of the eclipse (i.e. the orbital separation and stellar radii) and the photometric amplitude (i.e. the inclination and relative sizes) of the system. However, we can compare the number of detected eclipsing binaries in the GC with the number of eclipsing binaries in local OB associations. We make the assumption that the binary E60 with an amplitude of $\Delta m_K\approx0.45$ and average photometric error of $\sigma_K\approx 0.04$ mag, represents the detection limit.  Among the initial photometric sample of 113 early-type stars, 70 stars had photometric errors smaller or equal to E60. Only one star showed greater photometric variability on short timescales and that is IRS\,16SW. Some of the other stars showed variability on a $\sim0.1$ mag level with no obvious periodicity. Thus, out of the 70 stars, only two, IRS16\,SW and E60, are confirmed eclipsing binaries i.e. $F_{\rm EB}=3\pm2\%$ (at 1\,$\sigma$ confidence) with photometric amplitudes ($>0.4$ mag). 
The fraction of eclipsing binaries in local OB associations with similar amplitude variations $\Delta m \ge 0.4$ is $1.1\pm0.3$\% \citep[out of $\sim2400$\, OB stars, see][]{lefevre09}. \\
It is likely that the deduced fraction of massive eclipsing binaries in the GC reflects their initial formation fraction, given the very recent formation ($\sim$ 6\,Myr) of the massive stars, and given the strong observational bias of eclipsing binaries to be tight, dynamically hard binaries (e.g. the soft$/$hard boundary for E60 would be $a\approx300\, \rm R_{\odot}$; Section\,\ref{sec:binary_evol}), whereas eclipsing binaries have typically  $a\approx {\rm few}~ 10\, \rm R_{\odot}$. Therefore, while the large Poisson errors do not allow us to place tight constraints on the eclipsing binary fraction in the GC, we conclude that their initial fraction is close to the local value.

\section{Binary evolution in the Galactic Center}\label{sec:binary_evol}
\begin{figure}
\noindent \begin{centering}
\includegraphics[width=0.9\columnwidth]{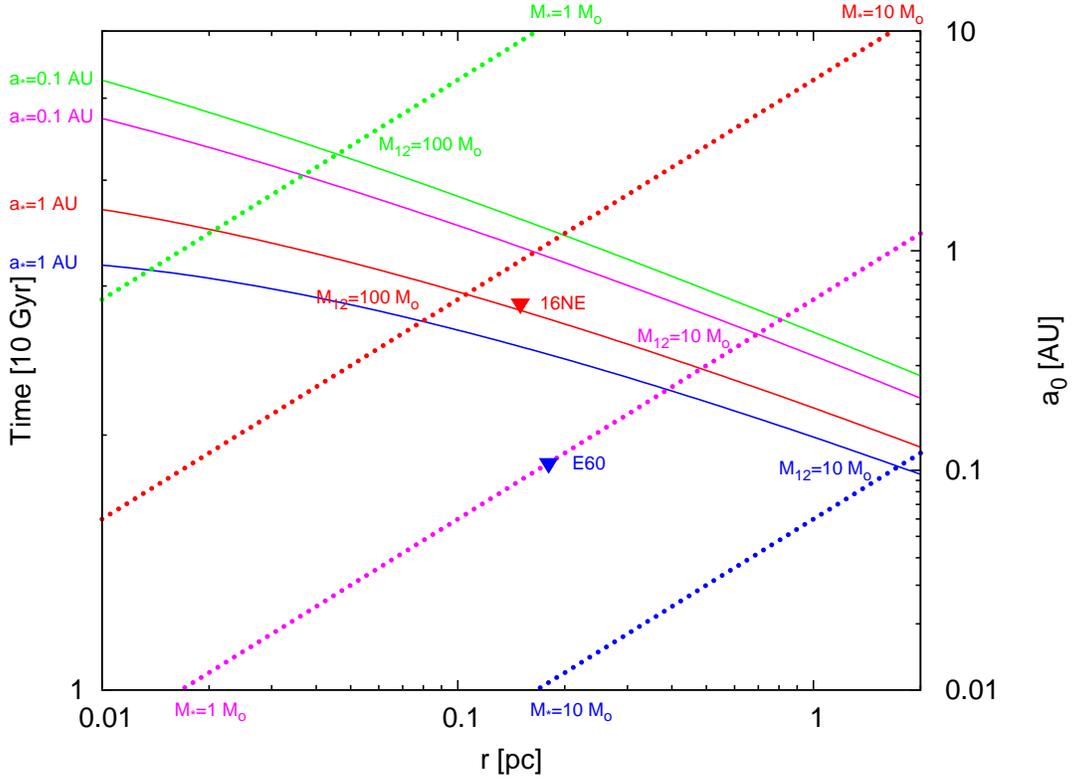}
\par\end{centering}
\caption{\label{f:binevol}The tidal separation timescale $T_{t}$ and the
soft/hard critical semi-major axis $a_{0}$ as function of distance
for the MBH in the GC, for very massive binaries $(M_{12}=100\, M_{\odot}$)
and moderately massive binaries $M_{12}=10\, M_{\odot}$). The tidal
separation timescales (solid lines) are evaluated for $a_{12}=0.1$
AU (of the order of that found for eclipsing binary E60), and $a_{12}=1$ AU (of
the order found for spectroscopic binary IRS16NE). The critical semi-major
axes (dotted lines) are evaluated for the assumed mean stellar mass
$\left\langle M_{\star}\right\rangle =10\, M_{\odot}$ expected very
close to the MBH due to mass segregation \citep[e.g.][]{alexander_hopman09},
or for a top heavy initial mass function (Alexander, Pfuhl \& Genzel,
2013, in prep.) and for $ $ $\left\langle M_{\star}\right\rangle =1\, M_{\odot}$,
as is expected further out for a universal initial mass function.
Approximate values for $a_{0}$ assuming $\left\langle M_{\star}\right\rangle =10\,
M_{\odot}$
are plotted for the binaries IRS16NE ($r\sim0.15$ pc $M_{12}\sim80\, M_{\odot}$
, $a_{12}\sin i\simeq1.2$ AU) and E60 ( $r\sim0.2$ pc $M_{12}\sim30\, M_{\odot}$,
$a_{12}\simeq0.1$ AU). These two binaries are close to their critical
semi-major axis. }
\end{figure}

The evolution of binaries in dense galactic nuclei can be strongly
modified by interactions with other stars and with the SMBH. We argue
here that such effects will not significantly influence the observed
properties and statistics of massive binaries in the GC.\\
Disruption by the Galactic SMBH affects only the very small fraction
of binaries that approach it within the tidal disruption radius, $r_{t}\simeq
a_{12}(M_{\bullet}/M_{12})^{1/3}$,
where $M_{12}=M_{1}+M_{2}$ is the binary's total mass, $a_{12}$
is its semi-major axis, and $M_{\bullet}$ is the SMBH mass. The timescale
for the center-of-mass of a binary at radius $r$ (assumed here to
be of the order of the semi-major axis of its orbit around the SMBH)
to be deflected by stochastic 2-body encounters with field stars to
an eccentric orbit that will lead to its tidal separation by the SMBH
is $T_{t}\sim\log(2\sqrt{r/r_{t}})T_{\mathrm{rlx}}(r)$, where $T_{\mathrm{rlx}}$
is the 2-body relaxation timescale, and the $\log$ term reflects
the typical time to diffuse in phase space into near radial orbits
with eccentricity $e_{t}=1-r_{t}/r$ \citep{1976MNRAS.176..633F}. The value of
$T_{\mathrm{rlx}}$ in the GC, and especially whether it is longer
or shorter than the age of the Milky Way (usually estimated by the
Hubble time, $t_{H}=10^{10}\,\mathrm{yr}$), depends on the yet unknown
dynamical state of the inner parsec, and in particular whether it
harbors a ``dark cusp'' of stellar remnants and faint stars \citep{alexander11}.
Estimates bracket it between
$T\mathrm{_{rlx}}\sim\mathrm{few\times10^{9}\,}\mathrm{yr}$
\citep{preto10} and $T\mathrm{_{rlx}}\sim\mathrm{few\times10^{10}\,}\mathrm{yr}$
\citep{merritt10}. Given that $T_{\mathrm{rlx}}\sim{\cal O}(t_{H})$,
and that typically $\log(2\sqrt{r/r_{t}})>1$, tidal separation of
binaries, especially short-lived massive ones, is negligible (Figure
\ref{f:binevol}). Rapid angular momentum relaxation by Resonant Relaxation
\citep{rauch96} is expected to become marginally relevant for this
tidal separation only in the inner $0.01$ pc \citep[Figure 7]{hopman06}.\\
The binary's internal orbit also evolves stochastically due to the
exchange of energy and angular momentum with field stars. The direction
of the energy exchange, that is, whether on average the binary gains
energy and becomes wider until it is disrupted (``evaporation''),
or whether it loses energy and shrinks until it coalesces, depends
on its softness parameter $s$, defined as the ratio between its binding
energy, $|E_{12}|=GM_{1}M_{2}/2a_{12}$, and the typical kinetic energy
of the field stars, $E_{K}\sim\left\langle M_{\star}\right\rangle \sigma^{2}$,
where $\sigma(r)$ is the 1D velocity dispersion. Soft binaries with
$s<1$ will ultimately evaporate, while hard binaries with $s>1$
will ultimately coalesce \citep{heggie75}. In terms of the binary's
semi-major axis, the soft/hard boundary is at the critical semi-major
axis $a_{0}=GM_{1}M_{2}/2\left\langle M_{\star}\right\rangle
\sigma^{2}\sim\left(M_{12}^{2}/8M_{\bullet}\left\langle M_{\star}\right\rangle
\right)r$,
where the approximations $\sigma^{2}\sim GM_{\bullet}/r$ (consistent
with the results of \citealt{trippe08}) and $M_{1}\sim M_{2}\sim M_{12}/2$
are assumed. Figure (\ref{f:binevol}) shows $a_{0}(r)$ for the very
massive binaries with $M_{12}\sim{\cal {O}}(100\, M_{\odot})$ and
the moderately massive binaries with $M_{12}\sim{\cal {O}}(10\, M_{\odot})$,
that are relevant for this study. IRS16NE and E60 are close to their
critical semi-major axis ($s\sim1$). It is therefore unclear what
direction their evolution would take, evaporation or coalescence.
However, the evolutionary timescales will in any case be much longer
than the binary's lifespan. A rough estimate of the time to coalescence
is $T_{c}\sim{\cal O}([s_{c}-s]T_{\mathrm{rlx}})$, where $s_{c}$
is the softness parameter where the binary's orbital decay is taken
over by non-dynamical effects (contact binary evolution or gravitational
wave losses), not considered here, while the time to evaporation is
$T_{e}\sim{\cal O}\left(\left[\left\langle M_{\star}\right\rangle
\left/M_{12}\right.\right]sT_{\mathrm{rlx}}\right)$
(\citealt{1987gady.book.....B}, Alexander, Pfuhl \& Genzel, 2013, in prep.). It
follows that as long as $0\ll s\ll s_{c}$, dynamical binary evolution
can be neglected for massive binaries in the GC.\\
These considerations do not apply for low-mass binaries, which are
expected to undergo substantial evolution in the inner $\sim0.1$ pc
of the GC \citep[e.g.][]{hopman09}. A detailed study of the dynamical
constraints that can be deduced from future detections of low-mass
binaries in the GC is presented in Alexander, Pfuhl \& Genzel (2013,
in prep.). 
\section{Discussion}
Our survey of more than a dozen massive OB/WR stars in the Galactic Center revealed two previously unknown binaries, the long period binary IRS\,16NE and the eclipsing binary E60. Within the uncertainties, the spectroscopic binary fraction $F_{\rm SB}=0.27^{+0.29}_{-0.19}$ in the GC seems to be close to the fraction observed in other dense clusters such as Trumpler 14 ($F_{SB}=0.14$). The same is true for the fraction of eclipsing binaries ($\Delta m \ge 0.4$) of $3\pm2\%$ compared to $1\%$ in other OB clusters. The fraction of multiple systems is significantly lower than unity. This is especially interesting since the multiplicity of stars formed in an SMBH accretion disk, is regulated by the cooling timescale of the parental disk \citep{alexander08}. Fast cooling timescales ($t_{\rm cool}<t_{\rm dyn}$), as are believed to be present in black hole disks \citep[e.g.][]{goodman03}, lead to binary fractions close to unity and mainly equal mass companions \citep{alexander08}. This is clearly not supported by our observations. The GC binary fraction can also not have been altered significantly by dynamical effects. Massive binary systems in the GC are either hard binaries or the evolution timescale exceeds the age of the OB/WR disk (6\,Myrs). With an extraordinary long period of 224 days, IRS\,16NE is an example for such a system that survived the dense cluster environment. The observed low binary fraction seems to be inconsistent with the current understanding of massive star formation in SMBH accretion disks. In that sense, the inferred binary fraction provides an additional constraint for future theoretical models that try to explain the formation of stars in the vicinity of SMBHs. 
\section{Conclusions}
\begin{itemize}
\item The massive 224 day long-period binary IRS\,16NE is a rare system, even in less extreme environments than the Galactic Center. Less than 10\% of all known OB binaries have longer periods. The high mass of the binary constituents allows the large separation even in the dense cluster environment. The binary is dynamically hard, and it is therefore expected to survive dynamical evaporation. 
% The binary is tightly bound and fulfills the binary hardness criterion. Tidal forces from the SMBH are negligible at the distance of the system. If such a large separation is stable in the dense stellar environment or should get more tightly bound during the lifetime of the system remains to be seen by future simulations.

\item We identified a new WR eclipsing binary (E60) at a distance of 4.7\arcsec~ from Sgr A*. The system is a contact binary with a short period of only 2.3 days and a system mass of $M_{\rm prim}\approx 20\,\msun$ and $M_{\rm sec}\approx 10\,\msun$. Together with IRS\,16SW, this star is the second known eclipsing binary in the Galactic Center. The system has a remarkably high velocity of $v_{\rm 3D}\approx 460\, \rm km/s$, which is close to or even higher than the escape velocity at the radial distance from Sgr A*. 

\item The spectroscopic binary fraction of the massive OB/WR stars in the Galactic Center is $F_{\rm SB}=0.27^{+0.29}_{-0.19}$, where the lower and upper limit represent the 95\% confidence interval. This result is broadly consistent with the massive binary fraction observed in dense young clusters (see discussion Sec.\,\ref{sec:observed_binary_fractions}). It seems to be inconsistent with the current understanding of star formation in SMBH accretion disks, which predicts a binary fraction close to unity. 
 
\item The eclipsing binary fraction ($\Delta m \ge 0.4$) in the GC is $3\pm2\%$. Within the errors this is consistent with the fraction in other dense OB clusters ($\approx1\%$). 

\end{itemize}

\acknowledgements{T.A. acknowledges support by ERC Starting Grant No. 202996,
DIP-BMBF Grant No. 71-0460-0101 and Israel Science Foundation I-CORE grant No. 1829/12. }

\bibliography{ms}

\end{document}